\documentclass[prb,aps,twocolumn,showpacs,superscriptaddress]{revtex4-2}
\usepackage{amsfonts,amsmath,multirow}

\usepackage[utf8]{inputenc}
\usepackage[T1]{fontenc}

\usepackage{makecell}
\usepackage{graphicx}
\usepackage{xcolor}
\usepackage{epstopdf}
\usepackage{dsfont}
\usepackage{physics}
\usepackage{placeins}
\usepackage{hyperref}
\usepackage{lipsum}
\usepackage{dcolumn}
\usepackage{latexsym}
\usepackage{orcidlink}
\usepackage{siunitx,amsmath}

\usepackage{mathtools}
\usepackage{hf-tikz}
\usepackage[normalem]{ulem}

\usepackage[]{newtxtext} 
\usepackage[nosymbolsc,smallerops,bigdelims]{newtxmath} 
\DeclareMathAlphabet{\mathcal}{OMS}{cmsy}{m}{n} 
\DeclareMathAlphabet{\mathbcal}{OMS}{cmsy}{b}{n} 

\usepackage{bm}

\newcommand{\kp}{k\! \vdot\! p}

\newcommand{\kk}{\bm{k}}

\newcommand{\eps}{\epsilon}

\DeclareUnicodeCharacter{0141}{\L{}}

\definecolor{bred}{HTML}{e31a1c}
\definecolor{bgreen}{HTML}{33a02c}
\definecolor{bblue}{HTML}{1f78b4}

\definecolor{armygreen}{rgb}{0.29, 0.33, 0.13}
\definecolor{newred}{RGB}{255,70,70}
\definecolor{newcyan}{RGB}{0,200,255}

\renewcommand{\cp}{{\mathrm{c.p.}}}

\newcolumntype{L}{D{.}{.}{3,4}}

\begin{document}
	
\title {Strain tunability of electron and hole g-factors in CsPb$X_3$: bulk and nanocrystals}

    \author{Aleksander Cząstkiewicz\orcidlink{0009-0000-0215-7181}}
    \affiliation{Department of Semiconductor Materials Engineering, Wroc\l aw  University of Science and Technology, Wybrze\.ze Wyspia\'nskiego 27, 50-370 Wroc{\l}aw, Poland}

    \author{Paweł Scharoch\orcidlink{0000-0001-7358-8755}}
    \affiliation{Department of Semiconductor Materials Engineering, Wroc\l aw  University of Science and Technology, Wybrze\.ze Wyspia\'nskiego 27, 50-370 Wroc{\l}aw, Poland}

    \author{Krzysztof Gawarecki\orcidlink{0000-0001-7400-2197}}
    \email{Krzysztof.Gawarecki@pwr.edu.pl}
    \affiliation{Institute of Theoretical Physics, Wybrze\.ze Wyspia\'nskiego 27, 50-370 Wroc{\l}aw, Poland}

    \begin{abstract}
         In this paper, we theoretically study the electron and hole $g$-factors in inorganic lead-halide perovskites for the reference cubic phase. We employ an 18-band $\kp$
        model and determine its parameters using target band structures obtained from DFT calculations. We use the linear response theory to calculate the $g$-factors and investigate their dependence on external strain. We then perform real-space calculations of the electron and hole $g$-factors in perovskite nanocrystals. We show that the bulk $g$-factors can be tuned over a wide range by external strain. In some
cases, strain can be used to tune the $g$-factor through zero.  By contrast, the carrier $g$-factors in CsPb$X_3$ nanocrystals are considerably less sensitive to strain due to quantum confinement.
         
\end{abstract}
	
	\maketitle
	
\section{Introduction}
\label{sec:intr}

Perovskites have attracted much attention owing to a broad range of applications in photovoltaics and optoelectronics~\cite{Even2015}. Their technological relevance also extends to perovskite lasers, offering low lasing thresholds and high cost‑effectiveness~\cite{Zhang2021}. Three-dimensional perovskites can also be utilized for the formation of nanocrystals (NCs). They enable applications such as lasers, photodetectors, LEDs, and even single‑photon emitters~\cite{Mi2025}. Such NCs are synthesized in various low‑cost material systems, including CsPb$X_3$. 
It is worth noting that lead halide perovskites exhibit pronounced spin-orbit coupling (SOC), primarily due to the heavy Pb atom. In combination with their strong optical response and prominent excitonic features, this large SOC provides a natural route to optical control of spin via spin‑selective excitonic transitions~\cite{Zhao2020}. In fact, experimental studies highlight the potential of MAPbI$_3$, from the point of view of optical spin injection and ultra-fast spin switches~\cite{Giovanni2015}. Further control of spin dynamics may be achieved by employing halide alloys of CsPb$X_3$ nanocrystals to realize phenomena such as spin mode locking~\cite{Kirstein2023}. Some reports also indicate a very long electron spin lifetime in perovskite nanocrystals~\cite{Belykh2026}. These diverse applications highlight the importance of developing advanced theoretical models to accurately predict the electronic, spin, and optical properties of perovskite materials and their nanostructures.

The g-factor is a figure of merit that determines the linear response of the system to the magnetic field, and 
knowing its actual values (for the electron and hole) is vital for engineering spin‑based quantum devices. In particular, there are two interesting regimes: An electron g-factor close to zero is plausible for architectures such as quantum repeaters~\cite{Kosaka2003}. Also, for low g-factor values, the system is resilient to magnetic noise~\cite{Burkard2023}, which can increase the coherence time. On the other hand, large g-factor values produce correspondingly large Zeeman splittings, which support reliable spin initialization and efficient transfer of quantum information~\cite{Vrijen2001}. In consequence, the g-factor tunability is of high interest to many applications. For regular III-V systems, it has been demonstrated that the g-factors can be tuned via strain~\cite{Charlebois1996,Nakaoka2005,Tholen2016}. Although carrier $g$-factors in halide perovskites have been investigated theoretically~\cite{Garcia-Arellano2022,Yu2016,Kirstein2022} and experimentally~\cite{Kirstein2022,Meliakov2023,Meliakov2024}, a systematic description of their tunability via external strain has not yet been fully established.

In conventional III–V bulk semiconductors such as InAs, InP, or GaAs, g‑factors are often evaluated using the Roth formula~\cite{Roth1959}, which follows from the eight‑band $\kp$ model. In this approach, their values are determined by the band gap $E_\mathrm{g}$, the Kane energy $E_\mathrm{p}$, and the spin-orbit splitting $\Delta$. For perovskites, however, it is known that electron g‑factor is strongly influenced by coupling to remote bands~\cite{Kirstein2022}. Therefore, the description beyond the eight-band $\kp$ is well motivated.

The g‑factor in confined systems differs significantly from the bulk case~\cite{Pryor2006,Meliakov2025}. Both experimental data and tight-binding calculations show that the carrier $g$-factors in lead-halide perovskite nanocrystals do not follow the same band-gap dependence as their bulk counterparts~\cite{Kirstein2022,Nestoklon2023}. 
The impact of the nanostructure size and shape can be intuitively viewed in terms of changes in spin‑correlated orbital currents~\cite{VanBree2016}. As a consequence, the real-space calculations are needed for nanostructures. 

In this paper, we theoretically study the properties of g-factors in lead halide inorganic perovskites CsPb$X_3$ (X=Cl, Br, I). We adopt the $18$-band $\kp$ description and combine it with Density Functional Theory (DFT) modeling to obtain the parameter sets. We explore how the g-factors can be tuned via external hydrostatic, biaxial, and uniaxial strain. We demonstrate that the values of the electron and hole g-factors change considerably in the presence of moderate strain. Finally, we calculate the electron and hole g-factors in perovskite NCs and demonstrate their tunability in the presence of external hydrostatic strain. We show that  CsPbCl$_3$ nanocrystals are considerably less sensitive to strain compared to the other systems due to the large band gap of the material.

The paper is organized as follows. In Section~II, we present the band structures obtained from the DFT calculations and the corresponding results of the 18-band $\kp$ model. Section~III is devoted to the bulk electron and hole $g$-factors. In Section~IV, we consider perovskite NCs and analyze the size and strain dependence of the Zeeman splitting. Finally, Section~V summarizes the main conclusions. The details related to the 18-band $\kp$ calculations are given in the Appendix.

\section{Band structures}

In this section, we present the DFT reference band structures and introduce the 18-band $\kp$ model used to describe the electronic states near the $R$ point.

\subsection{DFT calculations}

\begin{figure*}[tb]
		\begin{center}
			\includegraphics{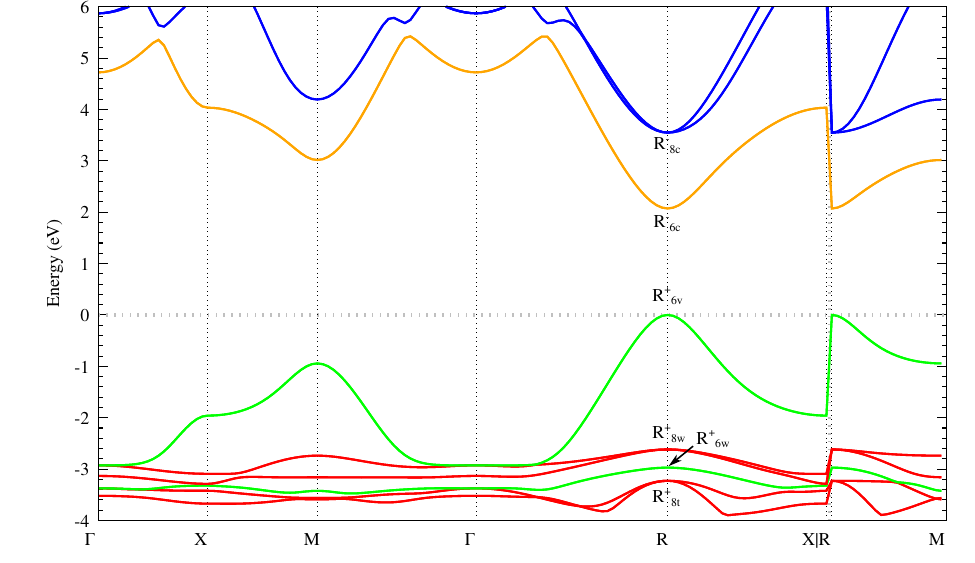}
		\end{center}
		\caption{\label{fig:bulk_bs} Band structure for CsPbBr$_3$. The colors of the bands indicate their irreducible representation at the $R$ point: red ($\Gamma^+_8$),  green ($\Gamma^+_6$), orange($\Gamma^-_6$), and blue ($\Gamma^-_8$).}
\end{figure*}

The reference band structures for the $\kp$ model parametrization were calculated within DFT using the Vienna $\emph{ab initio}$ Simulation Package (VASP)~\cite{VASPWebsite,VASPWiki,kresse1996efficient,kresse1999ultrasoft,kresse1996efficiency,kresse1993ab,kresse1994ab,kresse1994norm} software. Projector augmented-wave (PAW)~\cite{Blochl1994,kresse1999ultrasoft} potentials were employed with valence orbitals $5d^{10}6s^2 6p^2$, $5s^2 5p^6 6s^1$, $3s^2 3p^5$, $4s^2 4p^5$, and $5s^2 5p^5$ for Pb, Cs, Cl, Br, and I atoms, respectively. These calculations correspond to zero-temperature systems and neglect phonon effects.

As described in Ref.~\cite{Gawarecki2024a}, the rev-vdW-DF2~\cite{Klimes_2010,Klimes_2011,Lee2010} functional was used for geometry optimization to obtain an accurate atomic arrangement. The model takes into account the van der Waals interaction~\cite{Tao2017}.
The conjugate gradient algorithm was used for the optimization of the structure, following the specified convergence conditions. The energy difference required to reach was set to less than $5\cdot10^{-8}$~eV, and the convergence level of the norm of all forces was required to reach a value smaller than  $1.5\cdot 10^{-4}$~eV/\AA.
For $\bm{k}$ points sampling, we set the Monkhorst-Pack~\cite{Monkhorst1976} mesh of $6 \times 6 \times 6$ points.
The basis of plane-waves was defined by cut-off energy of 600 eV.
Following the methodology established in previous work~\cite{Gawarecki2024a}, the HSE06~\cite{Krukau2006} hybrid-functional with mixing parameter (AEXX\cite{VASPWiki} in VASP) adjusted to 0.6 was used for band structure calculations.
Spin-orbit coupling was included~\cite{Even2013}, except in
auxiliary simulations used for the symmetry analysis
(single group irreducible representations in Table~\ref{tab:symmetry}).

The exemplary band structure calculated for CsPbBr$_3$ is shown in Fig.~\ref{fig:bulk_bs}. 
The DFT simulations correspond to the zero-temperature case, and the crystal was enforced to maintain cubic symmetry. As the most important physics takes place in the vicinity of the Fermi level, we focus our attention on the bands in this region. The obtained results reproduce well known properties of lead-halide perovskites, i.e. they exhibit an inverted band ordering compared with conventional III--V semiconductors. Therefore, the lowest conduction bands consist of the spin-orbit split-off band as well as light- and heavy-electron branches.
The results correctly reproduce the direct character of the band gap at the $R$ point. 
The obtained $E_g = 2.072$~eV is close to the range of values of $2.29$ -- $2.36$~eV reported in the experimental works~\cite{Straus2022,Mannino2020,Heidrich1981,Tao2019}.

\subsection{The 18-band $\kp$ model}
\label{sec:band_kp}

\begin{table}
		\caption{\label{tab:symmetry} Irreducible representations assigned to the bands under consideration.\\[-0.07in] }
		\begin{ruledtabular}
			\begin{tabular}{llll}
				\makecell{Band block \\ (with SO)} & \makecell{Double \\ group irr.} & \makecell{Band block \\(without SO)} & \makecell{Single \\ group irr.}\\[0.1in]
				\hline\\[-0.1in]
				$R^{-}_{8c}$ & $\Gamma^{-}_{8}$ &  \multirow{2}{*}{$R^{-}_{4c}$} & \multirow{2}{*}{$\Gamma^{-}_4$} \\[0.02in]
                    $R^{-}_{6c}$ & $\Gamma^{-}_{6}$ &   & \\[0.02in]
                    \hline \\[-0.1in]
                    $R^{+}_{6v}$ & $\Gamma^{+}_{6}$ &  $R^{+}_{1v}$ &  $\Gamma^{+}_1$  \\[0.02in]
                    \hline \\[-0.1in]
                    $R^{+}_{8w}$ & $\Gamma^{+}_{8}$ &   \multirow{2}{*}{$R^{+}_{4w}$} &   \multirow{2}{*}{$\Gamma^{+}_4$}  \\[0.02in]
                    $R^{+}_{6w}$ & $\Gamma^{+}_{6}$ &   &    \\[0.02in]
                    \hline \\[-0.1in]
                    $R^{+}_{8t}$ & $\Gamma^{+}_{8}$ &  $R^{+}_{3t}$ &  $\Gamma^{+}_3$  \\[0.02in]
			\end{tabular}
		\end{ruledtabular}
	\end{table}

An 18-band $\kp$ model has previously been employed for III--V semiconductors and their heterostructures~\cite{Prevot2001,Vinter2002}. Here, we use an adapted version for perovskites. We take into account the 6 lowest-energy conduction bands, grouped into two blocks, $R^{-}_{8c}$ and $R^{-}_{6c}$, and the 12 highest-energy valence bands, grouped into the $R^{+}_{6v}$, $R^{+}_{8w}$, $R^{+}_{6w}$, and $R^{+}_{8t}$ blocks. Here, the superscript $\pm$ denotes the parity, while the letter in the subscript is used to distinguish the different band manifolds. Since, for the cubic space group Pm$\bar{3}$m, the little co-group at the $R$ point is isomorphic to $\mathrm{O}_\mathrm{h}$, the bands under consideration are classified according to the $\Gamma^{+}_{6}$, $\Gamma^{+}_{8}$, $\Gamma^{-}_{6}$, and $\Gamma^{-}_{8}$ irreducible representations (IRRs). However, the double-group irreducible representations alone do not uniquely identify the symmetry of the orbital part of the basis states,
as they describe the combined transformation of the orbital and spin degrees of freedom. For example, the same $\Gamma^{+}_{8}$ representation can originate from  $\Gamma^{+}_4 \otimes \Gamma^{+}_6 = \Gamma^{+}_8 \oplus \Gamma^{+}_6$, but also from $\Gamma^{+}_3 \otimes \Gamma^{+}_6 = \Gamma^{+}_8$. 
To gain further insight, we performed the symmetry analysis of the DFT bands using the \texttt{irrep} code~\cite{tsirkin_irrep,iraola2022irrep,elcoro2017double}. The assigned irreducible representations are listed in Table.~\ref{tab:symmetry}.
For CsPbI$_3$, the two lowest valence-band blocks are interchanged relative to the other compounds. Thus, from the highest to the lowest energy, the blocks are
ordered as: $R^{-}_{8c}$, $R^{-}_{6c}$, $R^{+}_{6v}$, $R^{+}_{8w}$, $R^{+}_{8t}$, and $R^{+}_{6w}$.
One should note that in the calculations performed without the spin-orbit coupling, the spinless
counterpart of the $R^{+}_{8t}$ block transforms as the $\Gamma^{+}_3$
representation involving the basis functions $2z^2-x^2-y^2$ and $\sqrt{3} \qty(x^2-y^2)$.

The 18-band $\kp$ Hamiltonian can be written in a matrix-block form
\begin{equation}
    \label{eq:H_block}
    H=
    \left(
    \begin{array}{*{6}{c}}
        \tikzmarkin[disable rounded corners=false,color=black,fill=gray!5,line width=0.1mm]{8kp}(2.95,-0.93)(-0.06,0.31)  
        \textcolor{bblue}{H_{\mathrm{8c8c}}} & \textcolor{bblue}{H_{\mathrm{8c6c}}} & H_{\mathrm{8c6v}} & H_{\mathrm{8c8w}} & H_{\mathrm{8c6w}} & H_{\mathrm{8c8t}} \\
        \textcolor{bblue}{H_{\mathrm{6c8c}}} & \textcolor{bblue}{H_{\mathrm{6c6c}}} & H_{\mathrm{6c6v}} & H_{\mathrm{6c8w}} & H_{\mathrm{6c6w}} & H_{\mathrm{6c8t}} \\
        H_{\mathrm{6v8c}} & H_{\mathrm{6v6c}} & \textcolor{red}{H_{\mathrm{6v6v}}} & \textcolor{red}{H_{\mathrm{6v8w}}} & \textcolor{red}{H_{\mathrm{6v6w}}} & \textcolor{red}{H_{\mathrm{6v8t}}} \\
        H_{\mathrm{8w8c}} & H_{\mathrm{8w6c}} & \textcolor{red}{H_{\mathrm{8w6v}}} & \textcolor{red}{H_{\mathrm{8w8w}}} & \textcolor{red}{H_{\mathrm{8w6w}}} & \textcolor{red}{H_{\mathrm{8w8t}}} \\
        H_{\mathrm{6w8c}} & H_{\mathrm{6w6c}} & \textcolor{red}{H_{\mathrm{6w6v}}} & \textcolor{red}{H_{\mathrm{6w8w}}} & \textcolor{red}{H_{\mathrm{6w6w}}} & \textcolor{red}{H_{\mathrm{6w8t}}} \\
        H_{\mathrm{8t8c}} & H_{\mathrm{8t6c}} & \textcolor{red}{H_{\mathrm{8t6v}}} & \textcolor{red}{H_{\mathrm{8t8w}}} & \textcolor{red}{H_{\mathrm{8t6w}}} & \textcolor{red}{H_{\mathrm{8t8t}}}
    \end{array}
    \right),
\end{equation}
where the Hamiltonian blocks corresponding to the conduction (valence) band were marked by blue (red) color. The highlighted block corresponds to the eight-band $\kp$ model for perovskites~\cite{Ompong2020,Yu2016,Fan2018,Gawarecki2024a}. 

The full Hamiltonian of the crystal can be written as
$$H(\bm{k}) = H^{(\bm{k})} + H^\mathrm{(str)},$$ where the first part describes unstrained band edges and the kinetic terms, and the second one accounts for strain. It is also beneficial to express the $\kp$ Hamiltonian in the invariant expansion form, as it exhibits a clear structure and makes the symmetry properties of the model explicit~\cite{Trebin1979,Bir1974,Winkler2003}. The block representation of the $H^{(\bm{k})}$ part is
	\begingroup
	\allowdisplaybreaks
	\begin{align*}
			H^{(\bm{k})}_{\mathrm{8c8c}} =&  E_{\mathrm{8c}}  \; \mathbb{I}_4 + \frac{\hbar^2}{2m_0}  \Bigg \{\gamma'_1 k_x^2 \; \mathbb{I}_4  - 2 \gamma'_2 \qty(J^2_x - \frac{J^2}{3}) k^2_x  \\& - 4 \gamma'_3  \{J_x,J_y\} \{k_x,k_y\}  + \cp \Bigg \},\\
			H^{(\bm{k})}_{\mathrm{6c6c}} =&  \qty( E_{\mathrm{6c}} + \frac{\hbar^2}{2m_0} \gamma'_1 k^2 ) \; \mathbb{I}_2 ,\\
			H^{(\bm{k})}_{\mathrm{6v6v}} =&  \qty( E_{\mathrm{6v}} - \frac{\hbar^2}{2m'_v} k^2) \; \mathbb{I}_2 ,\\
                H^{(\bm{k})}_{\mathrm{8w8w}} =& \; E_{\mathrm{8w}} \; \mathbb{I}_4 ,\\
                H^{(\bm{k})}_{\mathrm{6w6w}} =& \; E_{\mathrm{6w}} \; \mathbb{I}_2 ,\\
                H^{(\bm{k})}_{\mathrm{8t8t}} =& \; E_{\mathrm{8t}} \; \mathbb{I}_4 ,\\
			H^{(\bm{k})}_{\mathrm{8c6c}} =& -\frac{\hbar^2}{2m_0}  \Big \{6 \gamma'_2 T^\dagger_{xx} k^2_x + 12  \gamma'_3 T^\dagger_{xy} \{k_x,k_y\} + \cp \Big \},\\
                H^{(\bm{k})}_{\mathrm{8c6v}} =& \sqrt{3} P_0 \qty(T^\dagger_x k_x + \cp) ,\\
                H^{(\bm{k})}_{\mathrm{8c8w}} =& -\frac{2}{3} Q_0 \qty( \{J_y,J_z\} k_x + \cp),\\
                H^{(\bm{k})}_{\mathrm{8c6w}} =& -2 Q_0 \qty( T^\dagger_{yz} k_x+\cp ), \\
                H^{(\bm{k})}_{\mathrm{8c8t}} =& - \sqrt{30} \, R_0 \qty( D_x k_x + \cp), \\
                H^{(\bm{k})}_{\mathrm{6c6v}} =& -\frac{1}{\sqrt{3}}  P_0 (\sigma_x k_x + \cp), \\
                H^{(\bm{k})}_{\mathrm{6c8w}} =& -2 Q_0 (T_{yz} k_x+\cp), \\
                H^{(\bm{k})}_{\mathrm{6c8t}} =& \sqrt{6} \, R_0 (T_x k_x + \cp), 
		\end{align*}
        \endgroup
where $E_{6\alpha}$, $E_{8\alpha}$ are the band edges at the R point. The $m'_\mathrm{v}$ represents the remote band contributions to the effective mass in the $R^{+}_{6v}$ band; $\gamma'_1$, $\gamma'_2$, $\gamma'_3$ are the reduced Luttinger-like parameters; $P_0$, $Q_0$, and $R_0$ are parameters proportional to (interband) momentum matrix elements.  The ``$\cp$'' denotes cyclic permutation of indices. $J_i$ and $D_i$ are $4\times 4$ matrices, the $T_i$ are $2\times 4$ matrices connecting blocks, $T_{ij} = T_i J_j + T_j J_i  $, and $\mathbb{I}_n$ are unit matrices with the size of $n\times n$. The explicit definition of the matrices are given in Refs.~\cite{Trebin1979,Voon2009,Gawarecki2022}. The spin-orbit coupling parameters can be extracted from the energy differences $\Delta_\mathrm{c} = E_\mathrm{8c} - E_\mathrm{6c}$ and $\Delta_\mathrm{w} = E_\mathrm{8w} - E_\mathrm{6w}$.
We take into account the remote-band contributions only in the blocks corresponding to the 8-band model. The $k$-dependent terms in the ``8w'', ``6w'', and ``8t'' band
blocks are omitted.  Also, the interblock spin-orbit coupling matrix elements are neglected. Therefore, the following Hamiltonian blocks are set to zero: $H^{(\bm{k})}_{\mathrm{6c6w}}$, $H^{(\bm{k})}_{\mathrm{6v8w}}$, $H^{(\bm{k})}_{\mathrm{6v6w}}$, $H^{(\bm{k})}_{\mathrm{6v8t}}$, $H^{(\bm{k})}_{\mathrm{8w6w}}$, $H^{(\bm{k})}_{\mathrm{8w8t}}$, and
$H^{(\bm{k})}_{\mathrm{6w8t}}$.  The non-listed blocks are given by hermitian conjugates. 

\begin{table}[t]
\centering
\caption{The values of parameters for the 18-band $\kp$ model. The reference experimental band-gap values were selected from the
available literature~\cite{Zhang2018,Pandey2019,Yunakova2014,Yang2018,Tao2019,Straus2022,Mannino2020,Heidrich1981,Lingyan2020,Taiyang2017,Frolova2017,Dastidar2017}. }

\begin{ruledtabular}
\begin{tabular}{llll}
\label{tab:params}
& CsPbCl$_{3}$ & CsPbBr$_{3}$ & CsPbI$_{3}$ \\
\hline\\[-0.11in]
$a_\mathrm{0}$\hfill(\AA) &5.734   &5.911    &6.288  \\
\hline\\[-0.11in]
$E_\mathrm{8c}$\hfill(eV) &4.437   &3.549    &2.907  \\
$E_\mathrm{6c}$\hfill(eV) &2.991   &2.072    &1.412  \\
$E_\mathrm{6v}$\hfill(eV) &0.0   &0.0    & 0.0 \\
$E_\mathrm{8w}$\hfill(eV) &-2.999   & -2.618   &-1.666  \\
$E_\mathrm{6w}$\hfill(eV) &-3.074   & -2.972   &-2.445  \\
$E_\mathrm{8t}$\hfill(eV) &-3.469   &-3.228    & -2.425 \\
$P_0$   \hfill(eV$\,$\AA)  &7.409 &7.716 & 8.135  \\
$Q_0$   \hfill(eV$\,$\AA)  &3.606 &3.996 & 6.825   \\
$R_0$   \hfill(eV$\,$\AA)  &5.210 &5.406& 4.295  \\
$\gamma_1'$    &1.323   & 1.324   &0.611  \\
$\gamma_2'$    &0.197   &0.203    &0.445  \\
$\gamma_3'$    &-0.388   &-0.524    &-0.602  \\
$m_{v}'$        &0.565   &0.596   &0.643  \\[0.03in]
\hline\\[-0.11in]
$\Delta_\mathrm{c}$\hfill(eV) &1.446  & 1.477  &1.495 \\
$\Delta_\mathrm{w}$\hfill(eV) &0.075   &0.354   &0.779 \\[0.03in]
\hline\\[-0.11in]
$a_\mathrm{g}$\hfill(eV) &5.478   &5.767   &4.849 \\
$b_\mathrm{c}$\hfill(eV) &-2.316   &-2.354   &-2.350 \\
\hline\\[-0.11in]
$E^\mathrm{exp}_\mathrm{g}$\hfill(eV) &2.98   &2.29    &1.73  \\
\end{tabular}
\end{ruledtabular}
\end{table}

The strain-related part $H^\mathrm{({str})}$ has a form of the standard Bir-Pikus Hamiltonian, adapted to the band ordering that differs from
that of conventional III--V semiconductors~\cite{Bir1974,Winkler2003,Voon2009}. Here, we take into account the strain-dependence for $R^{-}_{8c}$, $R^{-}_{6c}$, and $R^{+}_{6v}$ bands, neglecting the others, which are separated by a larger energy distance from the bands of interest. 
Therefore, the Hamiltonian is given by
\begingroup
	\allowdisplaybreaks
	\begin{align*}
			H^\mathrm{(str)}_{\mathrm{8c8c}} =& \,   a_\mathrm{c} \qty(\eps_{xx} + \eps_{yy} + \eps_{zz}) \; \mathbb{I}_4  \\
                & - b_\mathrm{c} \qty[ \qty(J^2_x - \frac{J^2}{3}) \eps_{xx} + \cp ] \\
                & - d_\mathrm{c} \frac{2}{\sqrt{3}} \qty[  \{J_x,J_y\} \eps_{xy}  + \cp ],\\
			H^\mathrm{(str)}_{\mathrm{6c6c}} =& \, a_\mathrm{c} \qty(\eps_{xx} + \eps_{yy} + \eps_{zz}) \; \mathbb{I}_2,\\
                H^\mathrm{(str)}_{\mathrm{8c6c}} =& \, -3 b_\mathrm{c} \qty( T^\dagger_{xx} \eps_{xx} + \cp ) - 2 \sqrt{3} d_\mathrm{c} \qty(  T^\dagger_{xy} \eps_{xy} + \cp ), \\
			H^\mathrm{(str)}_{\mathrm{6v6v}} =& \, a_\mathrm{v} \qty(\eps_{xx} + \eps_{yy} + \eps_{zz}) \; \mathbb{I}_2 ,\\
            H^\mathrm{(str)}_{\mathrm{8c6v}} =& - 2 \sqrt{3} P_0 \qty(T^\dagger_x \sum_j \eps_{xj} k_j + \cp),\\ 
            H^\mathrm{(str)}_{\mathrm{6c6v}} =& \frac{2}{\sqrt{3}}  P_0  \qty( \sigma_x \sum_j \eps_{xj} k_j + \cp),
        \end{align*}    
\endgroup
where $a_\mathrm{c}$, $b_\mathrm{c}$, $d_\mathrm{c}$, and $a_\mathrm{v}$ are deformation potentials, and $\eps_{ij}$ are the strain tensor components.

\subsection{The $\kp$ model: numerical results}

\begin{figure*}[tb]
		\begin{center}
			\includegraphics[width=.9\textwidth]{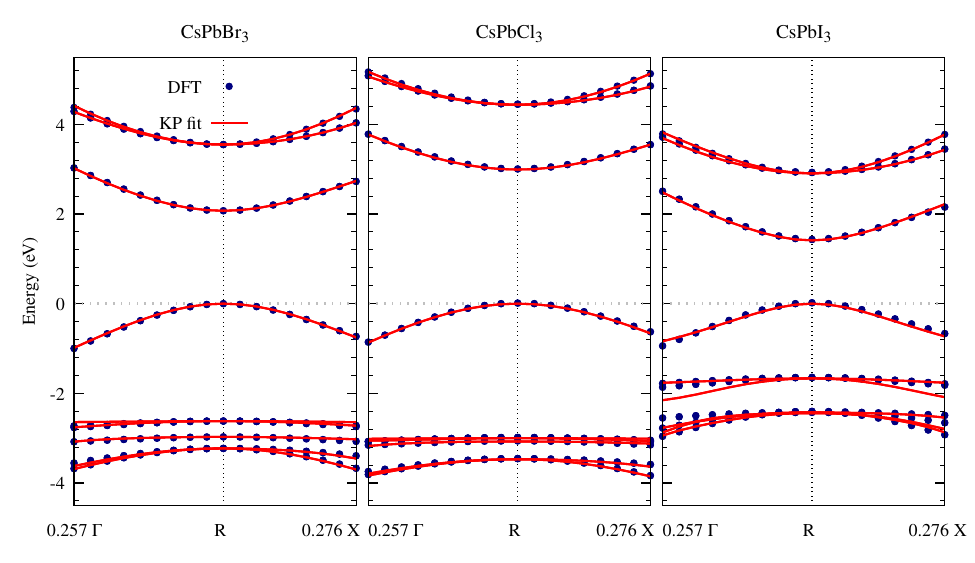}
		\end{center}
		\caption{\label{fig:kp_dft_bands} Band structures of CsPbBr$_3$, CsPbCl$_3$, and CsPbI$_3$ in the vicinity of the $R$ point,  calculated using the DFT and 18-band $\kp$ models. }
\end{figure*}

We found the $\kp$ parameters for CsPbBr$_3$, CsPbCl$_3$, and CsPbI$_3$ perovskites. First, the band edges and spin-orbit splittings $\Delta_\mathrm{c}$ and $\Delta_\mathrm{w}$
were extracted from the DFT results. Then, the values of $P_0$, $Q_0$, and $R_0$ were calculated from the momentum matrix elements using the VASP code, as described in Ref.~\cite{Gawarecki2022}. Finally, the remaining parameters accounting for the remote-band contributions $m'_\mathrm{v}$ and $\gamma'_{1-3}$ were obtained by fitting to the target DFT band structures. The values of all $\kp$ parameters are given in Table~\ref{tab:params}.
Figure~\ref{fig:kp_dft_bands} shows that the fitted model is in very good agreement with the DFT band structures in the vicinity of the $R$ point. For all considered materials, the splitting between the heavy- and light-holes (the $R^{+}_{8w}$ band manifold) is overestimated by the $\kp$ model. However, as these bands are not of interest, we do not attempt to refine their description by introducing an additional set of Luttinger-like parameters.

To determine the deformation potentials, we performed additional DFT calculations of the band-edges at the $R$ point as a function of external strain. We considered the hydrostatic ($\epsilon_{xx} = \epsilon_{yy} = \epsilon_{zz}$), biaxial ($\epsilon_{xx} = \epsilon_{yy} \neq 0$,  $\epsilon_{zz} = 0$), and uniaxial strain ($\epsilon_{xx} = \epsilon_{yy} = 0$,  $\epsilon_{zz} \neq 0$). 
For each strain configuration, the dimensions of the simulation cell were constructed from the equilibrium cubic lattice constant $a_0$ as
$a_i = a_0(1+\epsilon_{ii})$, with $i=x,y,z$.  The internal atomic coordinates were then relaxed using the rev-vdW-DF2 functional~\cite{Klimes_2010,Klimes_2011,Lee2010}, with the
deformed cell dimensions held fixed. The resulting geometries were used for the band-structure calculations at the $R$ point using the hybrid-functional HSE06~\cite{Krukau2006} with the same parameters as before.
However, in this type of calculation, the energy gauge is ill-defined~\cite{Kleinman1981}. In view of the lack of a proper energy reference, the results give information on the \textit{relative} changes only. Therefore,  we derive the band-gap deformation potential ${a}_\mathrm{g} = a_\mathrm{c} - a_\mathrm{v}$ instead of $a_\mathrm{c}$ and $a_\mathrm{v}$ separately.

\begin{figure*}[tb]
		\begin{center}
			\includegraphics[width=.9\textwidth]{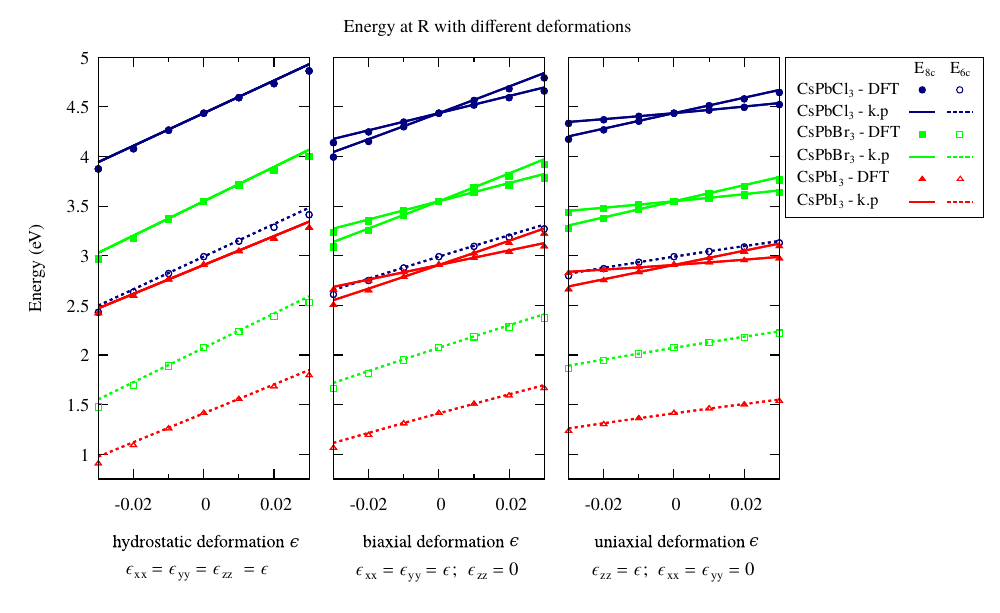}
		\end{center}
		\caption{\label{fig:kp_dft_strain} Band edges at the $R$ point in CsPbBr$_3$, CsPbCl$_3$, and CsPbI$_3$ perovskites as a function of strain, calculated using the DFT and the $\kp$ models. The filled symbols (solid lines) describe the $E_\mathrm{8c}$ band and the open symbols (dashed lines) correspond to the $E_\mathrm{6c}$ band.}
\end{figure*}
The band edges in CsPbX$_3$ perovskites calculated as a function of the hydrostatic, biaxial, and uniaxial strain are presented in Figs.~\ref{fig:kp_dft_strain}(a-c). Here, we set the $R^{+}_\mathrm{6v}$ band edge as the zero of energy, $E_\mathrm{6v}=0$. In this energy reference, $a_\mathrm{c} = a_\mathrm{g}$. As shown in Fig.~\ref{fig:kp_dft_strain}(a), under hydrostatic strain the
$E_{\mathrm{8c}}$ and $E_{\mathrm{6c}}$ band edges vary linearly (for DFT results: approximately linearly) for each material, with a common shift governed by the deformation potential $a_\mathrm{c}$. In the presence of biaxial [Fig.~\ref{fig:kp_dft_strain}(b)] and uniaxial [Fig.~\ref{fig:kp_dft_strain}(c)] strain, the $E_{\mathrm{8c}}$ manifold is split into light- and heavy-electrons. The DFT results are used as a target to fit the deformation potentials for the $\kp$ model. As shown in Fig.~\ref{fig:kp_dft_strain}, we obtain excellent agreement between the $\kp$ and DFT results. The fitted values of $a_\mathrm{g}$ and $b_\mathrm{c}$ are given in Table~\ref{tab:params}, where the sign of $b_\mathrm{c}$ was determined by inspecting the IRRs for the bands in the strained crystal. The results are in reasonable agreement with previous DFT studies~\cite{Wang2021}. Since the present analysis is restricted to normal strain components ($\epsilon_{xy}=\epsilon_{yz}=\epsilon_{zx}=0$), the deformation potential $d_\mathrm{c}$ does not enter the calculations and is therefore not determined.

Consistent with the literature~\cite{Wang2021}, the band gap decreases with negative hydrostatic strain (i.e., compression), which is opposite to the behavior typically observed in III--V material systems. As we consider perovskites with an enforced cubic phase, we obtain a monotonic strain dependence. However, for the orthorhombic phase, a nonmonotonic pressure dependence of the band gap has been reported ~\cite{Gong2022,Chen2024}. This behavior has been attributed to the interplay of two competing mechanisms: the change in volume and microscopic structural relaxation affecting bond lengths and bond angles~\cite{Chen2024}.

\section{The g-factors in bulk crystals}

\subsection{The theoretical model}

We model the electron and hole g-factors using linear response theory~\cite{Kiselev1998,Eissfeller2012,Gawarecki2020}, where the magnetic field is treated as a perturbation of the original Hamiltonian.  We first diagonalize the zero-field Hamiltonian $H(\mathbf{k})$ and
denote its cell-periodic Bloch eigenstates by $|u_{n\mathbf{k}}\rangle$. For each Kramers-degenerate band, these states span a two-dimensional subspace.  For a given direction of the
magnetic field, $\hat{\mathbf{B}}=\mathbf{B}/|\mathbf{B}|$, we diagonalize the Zeeman Hamiltonian within this subspace, obtaining the states $\ket{\psi_{+}(\mathbf{k})}$ and
$\ket{\psi_{-}(\mathbf{k})}$. The corresponding directional $g$-factor is then given by
\begin{align}
\label{eq:gfk}
    g(\kk) =& -\frac{1}{\mu_B} \Big ( \mel{\psi_+ (\kk)}{{\bm{\mu}(\kk) \cdot \hat{\mathbf{B}}}}{ \psi_+ (\kk)} \nonumber \\ & - \mel{\psi_- (\kk)}{{\bm{\mu}}(\kk) \cdot \hat{\mathbf{B}}}{\psi_- (\kk)} \Big ),
\end{align}
where $\mu_B$ is the Bohr magneton and the magnetic moment operator is given by
\begin{equation*}
    \bm{\mu} (\kk) = - \frac{\mu_B}{\hbar} \qty[ g_0 \bm{S}^\mathrm{(band)} + \bm{L}^\mathrm{(self)}(\kk)]
\end{equation*}
with $g_0 = 2$. The spin block matrices $S^\mathrm{(band)}_i$ are defined in the Appendix~\ref{app:details_spin}. The $\bm{L}^\mathrm{(self)}(\kk)$ part can be attributed to the angular momentum of self-rotating wave packet narrowly distributed around $\bm{k}$,
with the matrix elements for a given Bloch state $\ket{u_{n\kk}}$ given by~\cite{Chang2008} 
\begin{align*}
   \bm{L}^\mathrm{(self)}_n(\kk) & \equiv \mel{u_{n\kk}}{\bm{L}^\mathrm{(self)}(\kk)}{u_{n\kk}} \\ &= - i \frac{m_0}{\hbar} \sum_m \qty[E_n(\kk) - E_m(\kk)] \mathcal{\bm{A}}_{nm} (\kk) \times \bm{\mathcal{A}}_{mn} (\kk),
\end{align*}
where the summation index $m$ goes over states outside the selected two-dimensional subspace, and the vector-valued function 
\begin{equation*}
    \mathcal{A}_{nm} (\kk) = i \braket{u_{n\kk}}{\pdv{u_{m\kk}}{\bm{k}}}
\end{equation*}
is the Berry connection. With the relation~\cite{Xiao2010} 
\begin{align*}
    \braket{u_{n\kk}}{\pdv{u_{m\kk}}{\bm{k}}} = \frac{1}{E_m(\kk)-E_n(\kk)} \mel{u_{n\kk}}{\pdv{H}{\bm{k}}}{u_{m\kk}},
\end{align*}
the $\bm{L}^\mathrm{(self)}_n(\kk)$ components can be rewritten in form
\begin{align*}
   L^\mathrm{(self)}_{n,k}(\kk) & = - i \frac{m_0}{\hbar} \sum_{m\neq n} \sum_{i,j} \epsilon_{ijk}  \frac{\mel{u_{n\kk}}{\pdv{H}{k_i}}{u_{m\kk}} \mel{u_{m\kk}}{\pdv{H}{k_j}}{u_{n\kk}}}{E_n(\kk) - E_m(\kk)},
\end{align*}
where $\epsilon_{ijk}$ is the Levi-Civita tensor. When identifying the momentum and position matrix elements 
\begin{align*}
P_{i,nm}(\kk) &= \frac{m_0}{\hbar} \mel{u_{n\kk}}{\pdv{H}{k_i}}{u_{m\kk}}, \\
R_{i,nm}(\kk) &=  -\frac{i}{E_n(\kk) - E_m(\kk)} \mel{u_{n\kk}}{\pdv{H}{k_i}}{u_{m\kk}},
\end{align*}
one gets 
 \begin{align*}
   L^\mathrm{(self)}_{n,k}(\kk) & =  \sum_{m\neq n} \sum_{i,j} \epsilon_{ijk} R_{i,nm}(\kk) P_{j,mn}(\kk),
\end{align*}
which is a familiar form~\cite{Wozniak2020,Gawarecki2020}.

For the unstrained material, the values at the $R$ point are given by the following analytical expressions obtained within the perturbative
approach
\begin{align}
    g_\mathrm{h} &= 2 + \frac{4 m_0}{3 \hbar^2} P_0^2 \qty( \frac{1}{E_\mathrm{6v} - E_\mathrm{6c} } - \frac{1}{E_\mathrm{6v} - E_\mathrm{8c} } ), \label{eq:gh}\\
    g_\mathrm{e} &= -\frac{2}{3} + \frac{4 m_0}{3 \hbar^2} \Big ( \frac{P_0^2}{E_\mathrm{6c} - E_\mathrm{6v}} - \frac{Q_0^2}{E_\mathrm{6c} - E_\mathrm{8w}}   - \frac{2 R_0^2}{E_\mathrm{6c} - E_\mathrm{8t}} \Big ) \label{eq:ge}, 
\end{align}
where $g_\mathrm{h}$ is exactly the formula from Kirstein et al.~\cite{Kirstein2022}. 
%


\subsection{Results}

\begin{table*}[tb]
\begin{ruledtabular}
\begin{tabular}{lllllll}
\multirow{2}{*}{Compound} & \multicolumn{2}{c}{Literature values} & \multicolumn{2}{c}{Calculated values (DFT-derived parameters)} & \multicolumn{2}{c}{Calculated values (scissor shift) } \\ 
& Hole & Electron & Hole & Electron & Hole & Electron \\
\hline\\[-0.11in]
\multirow{2}{*}{CsPbBr$_3$}
& 0.65--0.85~\cite{Kirstein2022}
& 1.69--2.06~\cite{Kirstein2022}
& \multirow{2}{*}{-0.092}
& \multirow{2}{*}{1.836}
& \multirow{2}{*}{0.216}
& \multirow{2}{*}{1.460} \\

& 0.75~\cite{Belykh2019}
& 1.96~\cite{Belykh2019}
& & & & \\
\hline\\[-0.11in]
CsPbCl$_3$ & 1.14~\cite{Nestoklon2023} & 0.95~\cite{Nestoklon2023} & 0.953 & 0.694 & 0.947 & 0.703 \\
\hline\\[-0.11in]
CsPbI$_3$ & -0.33~\cite{Nestoklon2023} & 3.23~\cite{Nestoklon2023} & -2.218 & 3.204 & -1.103 & 2.073
\end{tabular}
\end{ruledtabular}
\caption{Calculated values of g-factors compared with experimental data for CsPbBr$_3$~\cite{Kirstein2022, Belykh2019}, and with other theoretical results for CsPbCl$_3$ and CsPbI$_3$~\cite{Nestoklon2023}.} 
\label{tab:gf}
\end{table*}

First, we calculated the electron and hole $g$-factors for unstrained bulk crystals at the $R$ point using the original DFT-derived parameter set employed for the band-structure calculations (as in Fig.~\ref{fig:kp_dft_bands}). The resulting values
are listed in Table~\ref{tab:gf} under ``DFT-derived parameters''. 
For CsPbBr$_3$, the calculated values can be compared with experimental results~\cite{Kirstein2022,Belykh2019}, where in the case of Kirstein et. al~\cite{Kirstein2022} the g-factors were determined as a function of angle. For reference, we also list the theoretical values from Nestoklon et. al~\cite{Nestoklon2023}. Good agreement is obtained for the electron $g$-factors in CsPbBr$_3$ and CsPbI$_3$, as well as for both carriers in CsPbCl$_3$. Larger discrepancies are found for the hole $g$-factor in CsPbBr$_3$ and CsPbI$_3$. It should be emphasized that our calculations assume an ideal cubic structure, whereas the reference experimental values~\cite{Kirstein2022} correspond to a low-temperature phase of lower symmetry.

To examine to what extent the discrepancies originate from the difference in the band-gap energy, we additionally considered a band-gap-corrected parameter set. We applied a scissor shift by setting
$E_\mathrm{6c} = E_\mathrm{g}^{\mathrm{exp}}$, with $E_\mathrm{6v}=0$, and simultaneously shifting the $E_\mathrm{8c}$ band-edge according to
\begin{align}
    E_\mathrm{8c} = E_\mathrm{6c} + \Delta_\mathrm{c} ,
\end{align}
while retaining the DFT-derived spin-orbit splitting $\Delta_c$ and all remaining parameters. The corresponding $g$-factors are listed in Table~\ref{tab:gf} under ``scissor shift''. The correction has only a
minor effect for CsPbCl$_3$. For CsPbBr$_3$, the correction improves the hole $g$-factor but deteriorates the agreement for the electron.

\begin{figure*}[tb]
		\begin{center}
			\includegraphics[width=.9\textwidth]{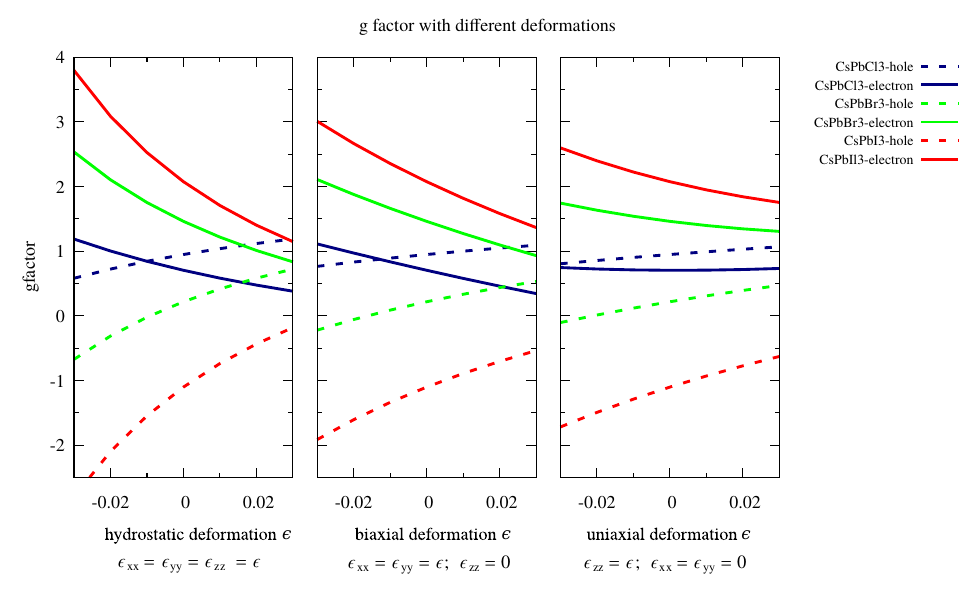}
		\end{center}
		\caption{\label{fig:gfactors_strain} Electron and hole g-factors at the $R$ point for CsPbBr$_3$, CsPbCl$_3$, and CsPbI$_3$,  calculated as a function of hydrostatic (a), biaxial (b), and uniaxial (c) strain, using the 18-band $\kp$ model. }
\end{figure*}
We next calculated the dependence of the electron and hole $g$-factors on external strain using the 18-band $\kp$ model with the deformation potentials introduced in Sec.~\ref{sec:band_kp}. Strain modifies the relative positions and dispersions of the bands, which in the presence of the spin-orbit coupling has an impact also on the spin properties. With hydrostatic strain [Fig.~\ref{fig:gfactors_strain}(a)], the electron $g$-factor decreases under tensile strain and increases under compressive strain.  The hole $g$-factor exhibits the opposite trend and increases under tensile strain. For both carriers, the observed dependence is pronounced, demonstrating that moderate lattice deformation can provide substantial control. We obtained the largest slopes for g-factors in CsPbI$_3$, which results from the smallest band-gap relative to the strain-related contributions. For CsPbBr$_3$, there is a change in the sign of $g_\mathrm{h}$, which means that the zero g-factor can be obtained by proper strain. 
The results for the biaxial strain ($\epsilon_{xx} = \epsilon_{yy}$,  $\epsilon_{zz} = 0$)  [Fig.~\ref{fig:gfactors_strain}(b)] and the uniaxial strain  ($\epsilon_{xx}=\epsilon_{yy}=0$, $\epsilon_{zz}\neq 0$)  [Fig.~\ref{fig:gfactors_strain}(c)] resemble dependencies obtained for hydrostatic strain. In this case, the slopes are modified by the contribution from $b_\mathrm{c}$ deformation potential, and the splitting in the $E_\mathrm{8c}$ manifold.

We note that considering the orthorhombic phase, as relevant for low-temperature experimental conditions, could modify the behavior presented here, as a nonmonotonic pressure dependence of the band gap
has been reported for this phase~\cite{Chen2024}. Such calculations, however, are beyond the scope of the present work.

\section{G-factors in perovskite nanocrystals}

\begin{figure}[tb]
		\begin{center}
			\includegraphics[width=.5\textwidth]{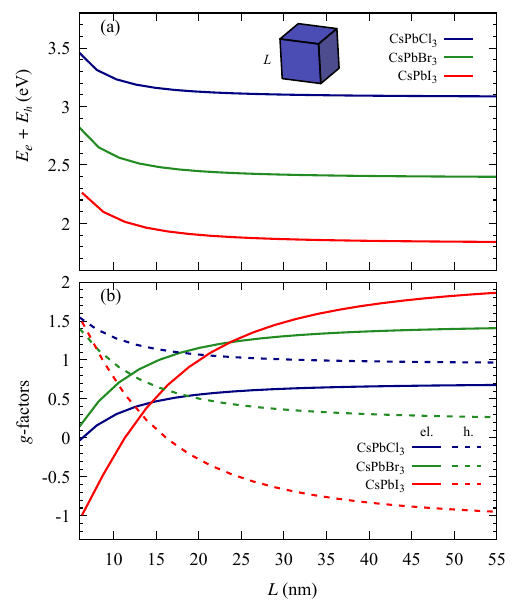}
		\end{center}
		\caption{\label{fig:gf_NC_L} (a) Electron-hole pair (lowest) energy as a function of NC size for CsPbBr$_3$, CsPbCl$_3$, and CsPbI$_3$. (b) Electron (solid lines) and hole (dashed lines) g-factors for the lowest pair of states in NCs.}
\end{figure}
To calculate the electron and hole $g$-factors in perovskite NCs, we performed real-space $\kp$ simulations within the envelope-function approximation. To this end, the $k_n$ components in the 18-band $\kp$ Hamiltonian were replaced by the corresponding derivatives, $k_n \rightarrow -i\,\partial/\partial x_n$. The details related to the implementation are given in the Appendix~\ref{app:details_18kp}. The magnetic field is introduced into the model through the spin-Zeeman term, as in the bulk Hamiltonian, and through gauge invariant phase factors in the discretization of the derivatives~\cite{Andlauer2008}. As in the previous section, we utilize the empirical band-gap correction (the scissor shift) of the $\kp$ parameters here. The hole-state energy is defined as the negative of the corresponding valence-band electron energy, $E_\mathrm{h}=-E_\mathrm{v}$.

The energy of the electron-hole pair ($E_\mathrm{e}+E_\mathrm{h}$ for the ground states of carriers) in CsPbX$_3$ nanocrystals, calculated as a function of the NC size $L$ (cube edge length), is shown in Fig.~\ref{fig:gf_NC_L}(a). As expected~\cite{Nestoklon2023}, the total energy increases significantly with decreasing NC size due to quantum confinement. The corresponding electron and hole $g$-factors are presented in Fig.~\ref{fig:gf_NC_L}~(b). The electron (hole) $g$-factor increases (decreases) with increasing NC size~\cite{Nestoklon2023}. For both carriers, the $g$-factor values approach their bulk limits as the NC size increases. In the case of electron CsPbCl$_3$  and both carriers in CsPbI$_3$, the $g$-factor crosses zero.

\begin{figure}[tb]
		\begin{center}
			\includegraphics[width=.5\textwidth]{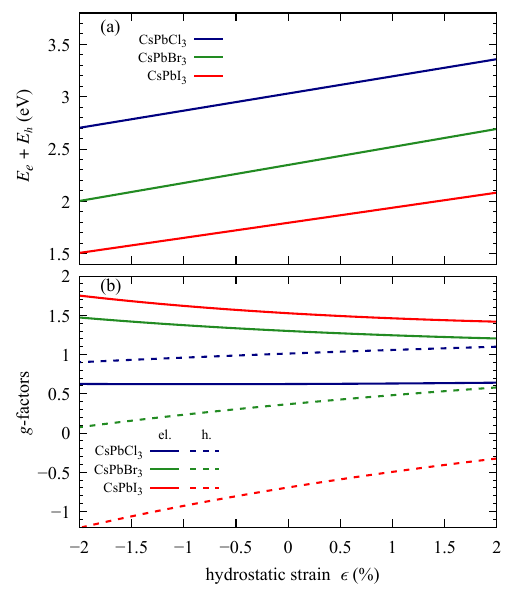}
		\end{center}
		\caption{\label{fig:gf_NC_eps} (a) Electron-hole pair (lowest) energy in a NC as a function of the external hydrostatic strain $\epsilon_{xx} = \epsilon_{yy} = \epsilon_{zz} = \epsilon$ for CsPbBr$_3$, CsPbCl$_3$, and CsPbI$_3$. (b) Electron (solid lines) and hole (dashed lines) g-factors for the lowest pair of states in NCs.}
\end{figure}
We calculated the electron-hole pair energy in the presence of hydrostatic strain [Fig.~\ref{fig:gf_NC_eps}(a)]. Here, we set the unstrained cube edge length to $L=50\,a_0$. The dependence is
approximately linear, with a slope governed by the deformation potential $a_\mathrm{g}$ (approximately as $3 a_\mathrm{g} \epsilon$). The corresponding electron and hole $g$-factors, again calculated from the lowest Zeeman doublets, are shown in Fig.~\ref{fig:gf_NC_eps}(b). As in the bulk case, the electron (hole) $g$-factor increases (decreases) under compressive strain. However, the strain dependence is less pronounced than in the bulk due to the confinement effect.


\section{Conclusions}
\label{sec:concl}

In conclusion, we have performed 18-band $\kp$ calculations for inorganic halide perovskite bulk crystals CsPbX$_3$ and their nanocrystals in the presence of external strain. Using DFT approach, we obtained target band structures, which were subsequently used to determine the $\kp$ parameters and deformation potentials.

Within this model, we calculated the electron and hole $g$-factors in reference cubic bulk crystals and analyzed their dependence on hydrostatic, biaxial, and uniaxial strain. We have shown that the carrier $g$-factors can be tuned by moderate strain. In particular, the electron and hole $g$-factors respond to strain in opposite ways: when the hole $g$-factor increases, the electron $g$-factor decreases. In some cases, this dependence can be used to tune the $g$-factor to zero, corresponding to a regime in which the system is more robust against magnetic noise. 

Finally, we modeled the electron and hole states in perovskite NCs and demonstrated that their $g$-factors can also be tuned by hydrostatic strain. However, the strain dependence is weaker in nanocrystals due to the effect of quantum confinement.

An interesting open question concerns the tunability of the $g$-factor by external pressure in the low-temperature orthorhombic phase, for which a nonmonotonic pressure dependence of the band gap has been
reported. However, exploring this regime would require a dedicated study.

\acknowledgments
Created using resources provided by Wroclaw Centre for Networking and Supercomputing (\url {http://wcss.pl}). We acknowledge the use of ChatGPT (OpenAI, GPT-5.6) for language
editing and for assistance in clarifying the presentation of selected scientific issues. All AI-assisted suggestions were critically reviewed and verified by the authors.
\FloatBarrier
\appendix
\section{Calculation details}

\subsection{The spin matrices for the 18-band $\kp$ model}
\label{app:details_spin}
The 18-band $\kp$ representation of the spin operator is given by
\begin{align*}
    S^\mathrm{(band)}_i &= \frac{1}{2}
    \begin{pmatrix} 
    \frac{2}{3} J_i & -2 T^\dagger_i & 0 & 0 & 0 & 0 \\
    -2 T_i & -\frac{1}{3} \sigma_i & 0 & 0 & 0 & 0 \\
    0 & 0 & \sigma_i & 0 & 0 & 0\\
    0 & 0 & 0 & \frac{2}{3} J_i & -2 T^\dagger_i & 0\\
    0 & 0 & 0 & -2 T_i & -\frac{1}{3} \sigma_i & 0 \\
    0 & 0 & 0 & 0 & 0 & \Xi_i
    \end{pmatrix},
\end{align*}
where

\begingroup
	\allowdisplaybreaks
\begin{align*}
    \Xi_x &= 
    \begin{pmatrix} 
    0 & 0 & 0 & -1 \\
    0 & 0 & -1 & 0 \\
    0 & -1 & 0 & 0 \\
    -1 & 0 & 0 & 0 
    \end{pmatrix}, \\
    \Xi_y &= 
    \begin{pmatrix} 
    0 & 0 & 0 & -i \\
    0 & 0 & i & 0 \\
    0 & -i & 0 & 0 \\
    i & 0 & 0 & 0 
    \end{pmatrix}, \\
    \Xi_z &= 
    \begin{pmatrix} 
    -1 & 0 & 0 & 0 \\
    0 & 1 & 0 & 0 \\
    0 & 0 & -1 & 0 \\
    0 & 0 & 0 & 1 
    \end{pmatrix} \\
\end{align*}
\endgroup

\subsection{18-band $\kp$ implementation for the NCs}
\label{app:details_18kp}

The NC simulations have been performed in the real space. The $k_i$ components in the Hamiltonian are replaced by the derivatives and discretized according the finite difference scheme. The details are described in the Appendix of Ref.~\cite{Gawarecki2018}. The magnetic field is introduced through the Zeeman term, and the transformation of the derivatives with additional phase factors~\cite{Andlauer2008}. Here, we utilize the Dirichlet boundary conditions and discretize the computational domain on a cubic grid, with the constant of $h_x = h_y = h_z = a_0/2$, where $a_0$ is the lattice constant of a given material. The $\kp$ simulations involve calculations on the domain from $10 \times 10 \times 10$ (the smallest NC) to $100 \times 100 \times 100$ (the largest NC). The resulting Hamiltonian matrices have sparse structure, and to diagonalize them, we have used SLEPC library~\cite{Hernandez2005,Romero2011}. To preserve the ellipticity, we set $\gamma'_1 = 1$, $\gamma'_2 = 0$, and $\gamma'_3 = 0$ for CsPbI$_3$.

\bibliography{references_KG,references_AC}

\end{document}